\documentclass[a4paper]{JHEP}

\usepackage{epsfig}

\newcommand{\apj}[3]{\emph{ApJ}\ {\bf #1} (#2) #3}

\newcommand{\astpar}[3]{\emph{Astropart. Phys.}\ {\bf #1} (#2) #3}

\def\gev{\;\rm GeV}

\title{Indirect detection of neutralino dark matter candidates
in anomaly-mediated supersymmetry breaking scenarios}

\author{Piero Ullio\\
        SISSA, via Beirut 4, 34014 Trieste, Italy \\
	E-mail:  \email{ullio@sissa.it}}

\abstract{
We consider a model for neutralino dark matter candidates arising in 
anomaly-mediated supersymmetry breaking schemes, and examine its testability 
through the search for exotic cosmic rays produced by neutralino pair 
annihilations in the dark halo of the Galaxy. We find that the model 
is already constrained by available antiprotons and positrons measurements 
and may be further tested in upcoming measurements of these cosmic ray 
species. We show also that the monochromatic gamma-ray flux from 
neutralino annihilations is enhanced in this model up to two orders of 
magnitude with respect to alternative scenarios. 
The gamma-ray flux detected by the Energetic Gamma Ray Experiment Telescope
in the direction of the Galactic center exceeds significantly the theoretical 
expectation of standard emission models. We prove that if at least 10\%
of this excess is due to gamma-ray radiation with continuum energy spectrum
from neutralino annihilations in the model under investigation,
the associated gamma-ray line will be detected by upcoming
gamma-ray experiments.}

\preprint{}


\begin{document}

\section{Introduction} 

The low energy phenomenology of a supersymmetric version 
of the standard model depends critically on the mechanism to describe 
supersymmetry (SUSY) breaking, still an open question. 
In a recently proposed scenario, SUSY is broken dynamically in a hidden 
sector with dominant SUSY breaking terms induced by 
anomalies~\cite{rs,giudiceetal}. 
Generic predictions of the anomaly mediated SUSY breaking (AMSB) 
scheme are the presence of a very heavy gravitino, with mass naturally
in the 100 TeV range or heavier, and the fact that
the lightest neutralino,  plausibly the lightest supersymmetric 
particle (LSP), can be either a Wino or a Higgsino. 

One of the byproducts of SUSY is the possibility for the LSP to be 
a dark matter candidate. From this point of view, the case of Wino- or 
Higgsino-like LSPs is generally regarded as not very attractive: 
Their thermal relic abundance is typically rather small; it is in 
the range compatible with the observed dark matter density only for 
very massive neutralinos, as heavy as 1~TeV 
for Higgsinos, see, e.g., Ref~\cite{eg}, and a few TeVs for Winos, see, 
e.g., Ref.~\cite{mny}. In the AMSB scenario, however, the nowadays density 
of the LSP is not fixed by its thermal relic density: An additional 
``non-thermal'' source is present due to decays into LSPs of gravitinos 
or moduli fields, fields that parameterize a flat direction of the theory 
and that dominate the energy density in the early Universe~\cite{ggw,mr}.
Fine-tuning in this mechanism is needed for the LSP to be the 
main dark matter component, certainly a drawback of the
AMSB scheme. On the other hand, we wish to stress here that the scenario 
with these LSPs accounting for the dark halo of our Galaxy has several 
phenomenological implications and will be probed by upcoming experiments 
for large regions in the SUSY parameter space.~\footnote{
The model has several more virtues; e.g., it solves the so-called 
``cosmological moduli problem'', see Ref.~\cite{mr}.}

There are several complementary techniques to identify dark matter 
neutralinos (for comprehensive reviews, see Refs.~\cite{jkg,lars}). 
Direct detection~\cite{dd} of AMSB candidates was discussed in 
Refs.~\cite{mr,mw}. We focus here on another technique:
the search for exotic cosmic rays, such as 
high-energy gamma-rays, antiprotons and positrons, produced by 
neutralino pair annihilations in the Galactic halo~\cite{cosmicrays}. 
This method looks particularly promising for neutralino dark matter 
candidates in the AMSB scenario, as both Winos and Higgsinos have a very 
high annihilation rate into gauge bosons, giving rise to strong sources 
of exotic cosmic rays. Indirect detection of Wino dark matter through 
antiproton and positron cosmic ray measurements was mentioned 
in Ref.~\cite{mr}; here, in a generic AMSB scenario, we derive the 
constraints on the model from available data and comment on the detection 
prospects for future experiments. Even more promising is indirect
detection with the next generation of $\gamma$-ray telescopes. We compute
the monochromatic gamma-ray flux~\cite{lines} from dark matter neutralino 
annihilations, a signal with no conceivable background from known 
astrophysical sources, and show that it is greatly enhanced for AMSB 
models. We compare with present data and en-light the cases in which such 
flux will be detected.

The outline of the paper is as follows: In Section~\ref{sec:model} 
we describe briefly the particle physics model we consider. 
In Section~\ref{sec:eppbar} we discuss implications of antiprotons 
and positron fluxes, while in Section~\ref{sec:gammas} we consider 
gamma-ray fluxes. Section~\ref{sec:concl} concludes.

\section{The MSSM in the AMSB scenario}
\label{sec:model}

We work in the minimal supersymmetric standard model (MSSM) as defined 
in Refs.~\cite{haberkane,jkg}; details on our notation are given in  
Ref.~\cite{eg}, while, for quantitative prediction, we use the {\sc DarkSUSY} 
computer code~\cite{DS}. We suppose the LSP is the 
lightest neutralino, defined as:
\begin{equation}
  \tilde{\chi}^0_1 = 
  N_{11} \tilde{B} + N_{12} \tilde{W}^3 + 
  N_{13} \tilde{H}^0_1 + N_{14} \tilde{H}^0_2\;.
\end{equation}
The coefficients $N_{1j}$, and hence the gaugino fraction
$Z_g = |N_{11}|^2 + |N_{12}|^2$, are obtained by 
diagonalizing the neutralino mass matrix; they are mainly a function of 
the bino and the wino mass parameters $M_1$ and $M_2$, and of the Higgsino 
parameter $\mu$. If $\left|\mu\right|$ is much smaller than 
$\left|M_1\right|$ and $\left|M_2\right|$, the lightest neutralino is 
Higgsino-like and $Z_g$ is close to zero. In the AMSB scenario the gaugino 
mass parameters are predicted in terms of the gravitino mass and their
ratio is set by the corresponding gauge coupling constants and
$\beta$-functions coefficients of the gauge coupling constants.
At the electroweak scale this relation becomes $M_1 \simeq 3\, M_2$, an
assumption we keep throughout the paper. If $\left|M_2\right|$ is much smaller 
than $\left|\mu\right|$, $Z_g$ is close to~1 and the lightest neutralino 
is Wino-like. We recall that, on the contrary, in alternative 
SUSY breaking scenarios, the GUT relation $M_1 \simeq 0.5\, M_2$ 
is usually assumed and a gaugino-like lightest neutralino is forced to be 
Bino-like.

For our purpose, it is not necessary to specify the whole mass spectrum
according to a specify AMSB model. The cosmic ray yield per LPS pair 
annihilation is fixed by the annihilation cross section and its branching
ratios. For both Winos and Higgsinos, the annihilation rate is fairly 
large (total rate $\sigma v \sim 10^{-24} \rm{cm}^3\,\rm{s}^{-1}$), 
and dominated by gauge boson final states.
Our predictions are then insensitive to most parameters in the sfermion 
and Higgs sectors, unless the annihilation cross section is enhanced further
in some special configuration. We make a sample parameter choice 
which removes this possibility: We assume that all sfermions are 
degenerate in mass, with masses 10~times the lightest neutralino mass 
and with no-mixing between right- and left-handed components. 
In the Higgs sector, we fix the pseudoscalar Higgs $A$ to be heavy,
$m_A = 500\gev$ and keep $\tan\beta$ as a free parameter.
Finally, to specify a model, rather than assigning $M_2$ and $\mu$, we
fix the lightest neutralino mass and gaugino fraction $M_{\chi}$ 
and $Z_g$, and the sign of $\mu$.

\section{Antiproton and positron fluxes}
\label{sec:eppbar}

Although antimatter seems to be scarce in the observed Universe,
a small flux of cosmic ray antiprotons and positrons is expected
from the interaction of primary cosmic rays with the interstellar
medium (see, e.g., Ref.~\cite{gaisserbook}). 
The measured $\bar{p}$ flux (\cite{bess98,caprice98}
and references therein) is compatible with the standard prediction 
for this secondary component, while strong exotic $\bar{p}$ sources
can be ruled out. An analogous conclusion follows from $e^+$ data
(\cite{heat95} and references therein).

The neutralino-induced source of exotic cosmic rays is not fully
specified once the particle physics model is chosen. The source 
function is proportional to the square of the neutralino density 
locally in space, a distribution which is only loosely known. 
In case of charged cosmic rays, the induced flux at earth is 
dominated by nearby sources; hence, for a given AMSB model, 
constraints on the local distribution of neutralinos can be derived.

We estimate the neutralino induced $\bar{p}$ flux according to 
the analysis in Ref.~\cite{pbar}, in which an accurate simulation 
of the antiproton yield per annihilation is performed and 
the propagation of antiprotons in the Galaxy is treated in 
a diffusive two-zone model. We suppose first that neutralinos are 
smoothly distributed in a spherical dark halo and compare
with the data collected by the {\sc Bess}~\cite{bess98} and 
{\sc Caprice}~\cite{caprice98} experiments in their latest flight
to derive the upper bound $\rho$ on the local neutralino density.
In Fig.~\ref{fig:fig1} we plot, with a solid line, a 
few isolevel curves for $\rho$ in the plane gaugino over higgsino 
fraction $Z_g/(1-Z_g)$ versus neutralino mass $M_{\chi}$ 
($\mu>0$ and $\tan\beta = 3$ are assumed; alternative choices give
hardly distinguishable curves). The displayed limits are at 90\% C.L. 
and are derived with a $\chi^2$ method under
a few conservative assumptions.
We compute the neutralino-induced $\bar{p}$ flux in all the energy
bins where data are available, but include in the statistical analysis
only those bins in which the expected flux exceeds the measured flux;
in such bins the secondary contribution is supposed to be subdominant
and is neglected. The values of $\rho$ displayed are derived assuming 
the dark halo is described by an isothermal sphere with a 5~kpc core radius;
a density profile singular towards the Galactic center would give more 
stringent limits~\cite{pbar}. 
\FIGURE[t]{
\epsfig{file=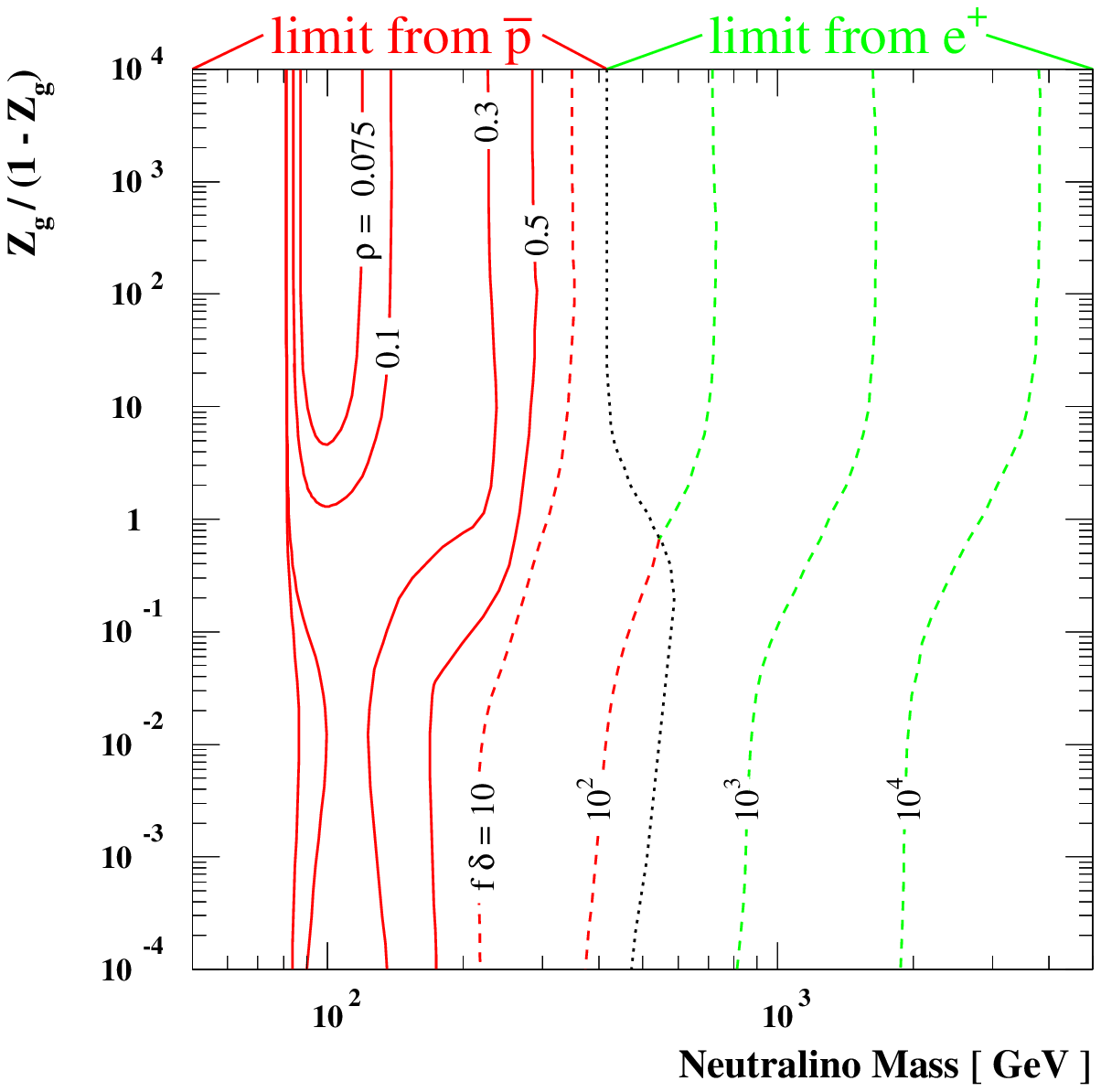,width=10.cm}
\caption{Isolevel curves for the maximal local halo density $\rho$
(solid lines) and maximal clumpiness parameter $f\,\delta$
(dashed lines), for neutralino dark matter candidates in the AMSB 
scenario, and in the plane gaugino over higgsino fraction versus
neutralino mass. Limits for lower neutralino masses (on the left-hand 
side with respect to the dotted line) are derived from antiproton 
cosmic ray data, while those for larger neutralino masses 
follow from positron cosmic ray data.}
\label{fig:fig1}
}
Finally, to model the diffusion of 
antiprotons in the Galaxy, we choose a standard set of the parameters in the 
propagation model compatible with measurements of other cosmic ray species 
(default choice in~\cite{pbar}); extreme values for these parameters may 
enhance (as well as suppress) the flux up to about 50\%. If one of these 
three assumption is changed, inducing a
scaling of the maximal allowed $\bar{p}$ flux
by roughly a factor $F$, the values assigned to the isolevel
curves should be scaled by $\sqrt{F}$.

Fig.~\ref{fig:fig1} shows that, for a given neutralino mass, Wino-like 
neutralinos (at the top in the figure) have more stringent limits than 
Higgsino-like (at the bottom) or mixed neutralinos. 
The lowest upper bound on the local density is about 0.06~GeV~cm$^{-3}$ 
for a pure Wino with mass about 100~GeV. There is a large region, with
$M_{\chi}$ between 80 and 200~GeV, in which $\rho$ is lower than the value
for the local dark matter density inferred from dynamical measurements
in the Galaxy, about 0.3--0.5~GeV~cm$^{-3}$ if the dark halo is assumed
to be spherical~\cite{tgg}; an additional dark matter component is then
needed.

N-body simulations of hierarchical clustering~\cite{moore} suggest
that dark matter may form substructures, or ``clumps'', rather than
being smoothly distributed in the halo (the picture we implemented 
so far). Larger local neutralino densities would enhance the sources
of exotic cosmic rays~\cite{ss}. Suppose that a fraction $f$ of the 
dark matter forms clumps with a some typical overdensity $\delta$,
defined as:
\begin{equation}
  \delta = \frac{1}{0.3\;\rm{GeV}\,\rm{cm}^{-3}}
  \frac{\int d^{\,3}r_{cl}\,
  \left(\rho_{cl}(\vec{r_{cl}})\right)^{2}}
  {\int d^{\,3}r_{cl}\,\rho_{cl}(\vec{r_{cl}})}\;,
\end{equation}
where the integration is over the extent of the clump and $\rho_{cl}$
is the neutralino density profile in the clump (see~\cite{clumpy} for 
further details). In this case, the induced cosmic-ray fluxes scale
with the product $f\,\delta$: comparing with the data, we have derived 
maximal values for this quantity. Isolevel curves for the maximal 
clumpiness parameter $f\delta$ are shown in Fig.~\ref{fig:fig1} as
dashed lines. For heavier and heavier 
neutralinos the limit inferred from the $\bar{p}$ flux becomes less 
and less stringent. We find that, for $M_{\chi}$ between 400 and 
600~GeV, on the right hand side with respect to the dotted curve 
plotted in $Z_g/(1-Z_g)$--$M_{\chi}$ plane, the analogous limit from 
$e^+$ cosmic ray measurements is more restrictive. This limit
is derived computing the $e^+$ flux according to the analysis in 
Ref.~\cite{be} and comparing it with data from the {\sc Heat} 
experiment~\cite{heat95}, with the same approach as for the $\bar{p}$ 
flux.

In the next years, space-based experiments~\cite{ams,pamela} are going 
to measure the $\bar{p}$ and $e^+$ fluxes with better statistics and much 
wider energy coverage. It will then be possible to set even more stringent
limits; furthermore, data at high energies may give evidence for the 
presence of this exotic component. As mentioned, AMSB models annihilate 
mainly into gauge bosons and these channels automatically
produce distinctive features
in the high energy spectra if the neutralino-induced fluxes are
at the level of the secondary components: a bump in the $e^+$ spectrum is 
expected at an energy roughly equal to $M_{\chi}/2$~\cite{kt,be},
a break in the $\bar{p}$ spectrum may appear at few tens of 
GeV~\cite{thesis}. In addition, a low energy exotic component in the
$\bar{p}$ flux might be identified~\cite{cosmicrays}.

\section{Gamma-ray flux}
\label{sec:gammas}

Gamma-rays with continuum energy spectrum are generated by the decay 
of $\pi^0$ mesons produced in jets from neutralino annihilations.
It is unfortunately a rather featureless flux, difficult to discriminate
from other plausible sources. When compared to the measured 
high-latitude gamma-ray background, it gives limits 
on the neutralino distribution in the AMSB scenario which are less restrictive
than the correspondent bounds we derived from $\bar{p}$ and $e^+$ 
measurements. 
A much better signature than the continuum contribution is given by 
the monochromatic $\gamma$-ray lines which arise from the 1-loop-induced 
neutralino annihilation into the $2 \gamma$ and $Z \gamma$ final states
(the energy of the $\gamma$-rays in the final state is equal,respectively, 
to $M_{\chi}$ and $M_{\chi} - M_Z^2/4\,M_{\chi}$).
Only recently, the amplitude of these processes was computed at full one 
loop level and for a generic MSSM~\cite{2g,zg}. Predictions for these two 
cross sections, in the AMSB scenario and as a function of neutralino mass,
are shown in Fig.~\ref{fig:fig2}.
We have chosen a few values of $Z_g$ (one for each shaded area as 
indicated in the figure) and varied $\tan\beta$ between 1 and 60 
(spread in each shaded area). The main contribution to the cross section
comes from diagrams with charginos and $W$ bosons in the internal loops;
these are sensitive to the mass splitting between the lightest neutralino 
and the lightest chargino, which is slightly modified by $\tan\beta$. 
\FIGURE[t]{
\epsfig{file=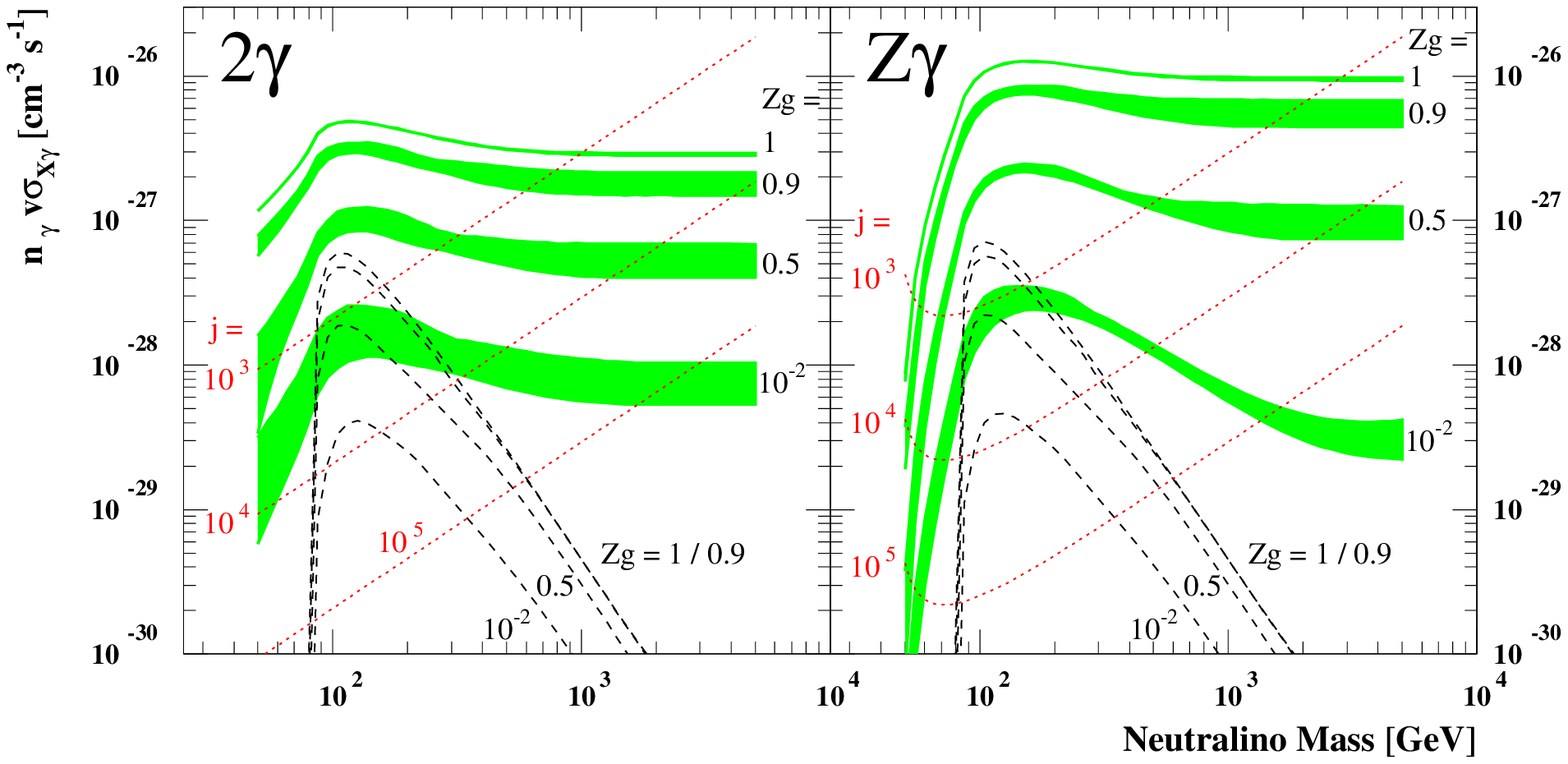,width=16.cm}
\caption{Annihilation rate into the two photon and the photon--$Z$ boson
final states for neutralinos in the AMSB scenario versus neutralino mass; 
a few values of the gaugino fraction $Z_g$ are considered. 
Also shown are 
$5\sigma$ sensitivity curves for a generic next generation air Cherenkov 
telescope~\protect\cite{magic,hess,veritas} observing the Galactic center:
Dotted lines correspond to a few sample dark matter distribution,
including a Moore et al. profile~\protect\cite{moore2} for $j=10^5$ and 
a NFW profile~\protect\cite{nfw} for $j=10^3$. Dashed lines are obtained, 
for each value of $Z_g$,
by fixing the neutralino distribution in such way that the
neutralino-induced $\gamma$-ray flux with continuum energy 
spectrum is compatible with the excess found in {\sc Egret} 
data~\protect\cite{mh}.} 
\label{fig:fig2}
}
In all cases the cross sections are fairly large; for Wino-like neutralinos 
they are a couple of orders of magnitude larger than the maximal values
in other SUSY breaking scenarios~\cite{bub}. The case $Z_g = 10^{-2}$ 
corresponds to a good accuracy to the lower limit for each of the two 
cross sections; this is again a peculiarity of the AMSB scenario, as,
on the other hand, there is essentially no lower bound to both cross 
sections for Bino-like dark matter candidates.
Note also that in most cases the annihilation rate into the 
$Z \gamma$ final state is larger than twice the annihilation
rate into two photons.

The $\gamma$-ray flux in a given direction of the sky is obtained
summing the contributions along the line of sight; the largest flux
is expected in the direction of the Galactic center, where an enhancement 
in the dark matter density is usually postulated. 
The distribution of dark matter in the inner part of the Galaxy is still 
a controversial issue. 
N-body simulations find singular profiles, scaling as $1/r$~\cite{nfw}
or $1/r^{1.5}$~\cite{moore2} as the galactocentric distance $r\rightarrow 0$.
These profiles 
correspond to snapshots of the Galaxy before the baryon infall; 
the appearance of the massive black hole at the Galactic center and of 
the stellar components may have sensibly modified such pictures with
further enhancements (but a depletion is possible as well) 
in the central dark matter density~\cite{gs,uzk}. 
There may be even the possibility that density of 
neutralinos in the galactic center region is substantial, while their 
contribution to the local dark matter density is subdominant.
We parameterize the dependence of the $\gamma$-ray flux
on the neutralino distribution through the dimensionless
quantity:
\begin{equation}
J\left(\psi\right) = \frac{1} {8.5\,\rm{kpc}} 
\int_{l.\,o.\,s.} d\,l(\psi)\;
\frac{\rho^2(l)}{(0.3\,\rm{GeV}\,\rm{cm}^{-3})^{2}}\;,
\end{equation}
where $\rho(l)$ is the neutralino density at the distance l
from the Earth along the line of sight. Then , we introduce
$j = \langle\,J\left(\psi=0\right)\rangle_{\Delta\Omega = 10^{-3} 
\rm{sr}}$, the average of $J$  over the field of view 
$\Delta\Omega = 10^{-3} \rm{sr}$
and in the Galactic center direction $\psi=0$.
E.g., the Moore et al. profile~\cite{moore2}, scaling as $1/r^{1.5}$ and 
normalized to $0.3\,\rm{GeV}\,\rm{cm}^{-3}$ at the Sun's galactocentric 
distance, gives $j$ equal to about $10^5$; for the Navarro Frenk and White
profile~\cite{nfw}, scaling as $1/r$ and with the same normalization,
$j \sim 10^3$.
For a given $j$, we can derive the sensitivity curve for a typical 
air Cherenkov gamma-ray telescope presently under 
development~\cite{magic,hess,veritas}: 
we consider an instrument with a $10^9$~cm$^2$ effective area and 
15\% energy resolution, assume a $10^6$~s exposure, and use a standard 
estimate for the background (see~\cite{bub} for details).
We plot the relative $5\sigma$ sensitivity curves for a few values of $j$
in Fig.~\ref{fig:fig2} as dotted lines; 
sensitivity curves for the upcoming Gamma-ray Large Area Space 
Telescope ({\sc Glast})~\cite{glast} are comparable.
As it can be seen, even for moderate values of $j$, large portion of the 
parameter space will be tested.

The Energetic Gamma Ray Experiment Telescope ({\sc Egret})
has mapped the diffuse Galactic gamma-ray flux up to an energy of about 
20~GeV~\cite{egret}. 
The flux detected in the direction of the Galactic center exceeds 
significantly the theoretical expectation of standard $\gamma$-ray 
emission models~\cite{mh}. One of the conceivable interpretation of the 
data is the presence of an exotic component from the annihilation of dark 
matter particles~\cite{mh}. If we invoke the case of AMSB models,
for given $M_{\chi}$ and $Z_g$, we can infer the maximal value of 
$j$ such that the neutralino-induced gamma-ray flux with continuum energy 
spectrum is consistent with or do not exceed the flux inferred
from the data analysis. The line and continuum energy fluxes scale 
according to the same factor $j$; we can now consider this upper limit,
say $j_{max}(M_{\chi},Z_g)$, derived from the continuum energy flux, and
compute the corresponding sensitivity curves for the monochromatic fluxes
(same assumptions as in the example above). These curves are shown in 
Fig.~\ref{fig:fig2}
as dashed lines, one for each value of $Z_g$.
Comparing each curve with the prediction
for the line cross sections, we can formulate a rather strong 
statement: If at least 10\% of the excess found by {\sc Egret} is due 
to the gamma-ray radiation with continuum energy spectrum from a 
neutralino dark matter candidate in a AMSB model, the associated 
gamma-ray lines will be detected by future gamma-ray experiments for 
any values of the SUSY parameters in the model.
Finally, by comparing the sensitivity curves obtained for a fixed $j$
with those obtained by fixing the continuum energy flux, we find that
some combination of SUSY parameters and dark matter distributions
are already inconsistent with gamma-ray measurements in the Galactic 
center direction. Consider one sensitivity curve for a given $j$ 
(dotted line) and one labelled by a value of $Z_g$ (dashed line): 
values of $M_{\chi}$ between the intersection 
points of the two curves are excluded (unless we invoke an efficient 
mechanism to absorb GeV gamma-rays in the Galactic center region).

\section{Conclusion}
\label{sec:concl}

We considered models for neutralino dark matter candidates arising in 
anomaly mediated supersymmetry breaking scenarios, and discussed their
testability through the search of exotic cosmic rays produced by 
neutralino annihilations in Galactic halo. We have shown that available
cosmic ray data already place significant constraints, while
even tighter limits are expected with data from upcoming experiments.
At the same time, new data, both on the antiproton and positron
cosmic ray flux as well as from gamma-ray surveys, may lead to the 
identification of these dark matter candidates. 
We proposed indirect detection through the search 
of a monochromatic $\gamma$-ray flux in the direction of the Galactic 
center. We showed that such flux may be observable if there is a moderate 
enhancement of the dark matter density in the Galactic center region,
and will be detected if the excess in the gamma-ray radiation from the 
Galactic center found in {\sc Egret} data is caused by self annihilations
of dark matter candidates in this class of models.

\acknowledgments
The author would like to thank Lars Bergstr{\"o}m and Joakim Edsj{\"o} 
for discussions.
This work was supported by the RTN project under grant
HPRN-CT-2000-00152.


{}

\end{document}